\documentclass[aps,prl,floats,twocolumn,showpacs,superscriptaddress,nobibnotes]{revtex4-2}
\usepackage[svgnames]{xcolor} 
\usepackage{graphicx,epsfig}
\usepackage{times}
\usepackage{amssymb,amsmath,multirow,rotate,color}
\usepackage{xcolor}
\usepackage[normalem]{ulem}

\definecolor{albicocca}{rgb}{0.98, 0.7, 0.2}
\definecolor{internationalorange}{rgb}{1.0, 0.31, 0.0}
\definecolor{giocolor}{RGB}{0, 150, 100}

\usepackage{float}
\usepackage{lipsum} 
\usepackage{ragged2e}
\usepackage{soul}
\DeclareMathAlphabet\mathbfcal{OMS}{cmsy}{b}{n}

\begin{document}


\title{Hyperedge overlap drives explosive collective behaviors in systems with higher-order interactions}

\author{Federico Malizia}
\thanks{These two authors contributed equally}
\affiliation{Department of Physics and Astronomy,  University of Catania, 95125 Catania, Italy}
\affiliation{Department of Condensed Matter Physics, University of Zaragoza, 50009 Zaragoza, Spain}
\author{Santiago Lamata-Otín}
\thanks{These two authors contributed equally}
\affiliation{Department of Condensed Matter Physics, University of Zaragoza, 50009 Zaragoza, Spain}
\affiliation{GOTHAM lab, Institute of Biocomputation and Physics of
Complex Systems (BIFI), University of Zaragoza, 50018 Zaragoza, Spain}
\author{Mattia Frasca}
\affiliation{Department of Electrical, Electronics and Computer Science Engineering, University of Catania, 95125 Catania, Italy}
\author{Vito Latora}
\affiliation{Department of Physics and Astronomy,  University of Catania, 95125 Catania, Italy}
\affiliation{School of Mathematical Sciences, Queen Mary University of London, London E1 4NS, United Kingdom}
\affiliation{Complexity Science Hub Vienna, A-1080 Vienna, Austria}

\author{Jes\'us G\'omez-Garde\~nes}
\affiliation{Department of Condensed Matter Physics, University of Zaragoza, 50009 Zaragoza, Spain}
\affiliation{GOTHAM lab, Institute of Biocomputation and Physics of
Complex Systems (BIFI), University of Zaragoza, 50018 Zaragoza, Spain}

\date{\today}


\begin{abstract}
Recent studies have shown that novel collective behaviors emerge in complex systems due to the presence of higher-order interactions. However, how the collective behavior of a system is influenced by the microscopic organization of its higher-order interactions remains still unexplored. In this Letter,  we introduce a way to quantify the overlap among the hyperedges of a higher-order network, and we show that real-world systems exhibit  different levels of hyperedge overlap. We then study models of complex contagion and synchronization of phase oscillators, finding that hyperedge overlap plays a universal role in determining the collective dynamics of very different systems. Our results demostrate that the presence of higher-order interactions alone does not guarantee abrupt transitions. Rather, explosivity and bistability require a microscopic organization of the structure with a low value of hyperedge overlap.  
\end{abstract}

\maketitle



\textit{Introduction.} 
In the last two decades network science has largely contributed to understanding how collective behaviours emerge in a complex system.  Representing and characterizing the intricate pattern of interactions among the constituents of a complex system as a graph ~\cite{newman,latorabook} has allowed to investigate how the system's structure affects its dynamics~\cite{barrat_book_2008}. 
A large variety of dynamical processes,  
ranging from percolation \cite{Percolation} to synchronization \cite{ARENAS200893}, epidemics \cite{RevModPhys.87.925} and social cooperation \cite{SZABO200797}, 
has been considered and explored, and  
this has led to the discovery of novel microscopic mechanisms to trigger and control collective behaviors.  

The advancement of data gathering techniques has pinpointed that pairwise interactions alone do not fully capture the interaction backbone of complex systems, prompting the need to explore the role of higher-order interactions~\cite{report_Ho,otra_review,bick2022higherorder}. In the recent years, network science has put the focus on the development of a general framework comprising group interactions for the study of collective behaviors well beyond the limitations of pairwise interactions~\cite{Iacopini,arenas_expl,yamir_hg,hosynch,exp2,arenas_synch,gambuzza2021stability,gallo2022synchronization,civilini2023explosive}. In particular, it has provided a natural pathway towards explosive transitions \cite{nature,bick}, with examples spanning social contagion \cite{Iacopini,arenas_expl,yamir_hg}, synchronization \cite{hosynch,exp2,arenas_synch} or game theory \cite{civilini2023explosive}. 
Nevertheless, little has been said about the role that the microscopic organization of higher-order  interactions of a system has on the onset of collective phenomena. 

In this Letter, we show that the presence of higher-order interactions alone is not sufficient to lead to explosive transitions. What is crucial is the way in which the nodes interact in groups, and how many nodes of a group are also present in other groups. To quantify this, we introduce a way to  measure the overlap among the hyperedges of a higher-order network. 
Akin to the clustering coefficient of a graph, our measure  
of hyperedge overlap evaluates the number of nodes shared among the different hyperedges of a hypergraph. 
We first study hypergraphs describing higher-order interactions in various real
complex systems, showing that they exhibit a large variety of values of hyperedge overlap. 
We then investigate if and how the different level of hyperedge overlap of a system affects the emergence and properties of its collective behavior. 
By focusing on two radically different dynamical processes, namely social contagion and synchronization of coupled dynamical systems, we highlight the universal effect generated by  hyperedge overlap in higher-order structures. In particular, we show that hypergraphs with low hyperedge overlap undergo explosive transitions, characterized by a bistable region where both an active/synchronized state and an absorbent/incoherent state coexist. Conversely, hypergraphs with a hyperedge overlap larger than a critical value can only exhibit continuous transitions. These results reveal that it is the structural organization of hyperedges what drives the way collective behaviors emergence in systems with higher-order interactions.

\begin{figure}[t!]
\includegraphics[width=1\linewidth]{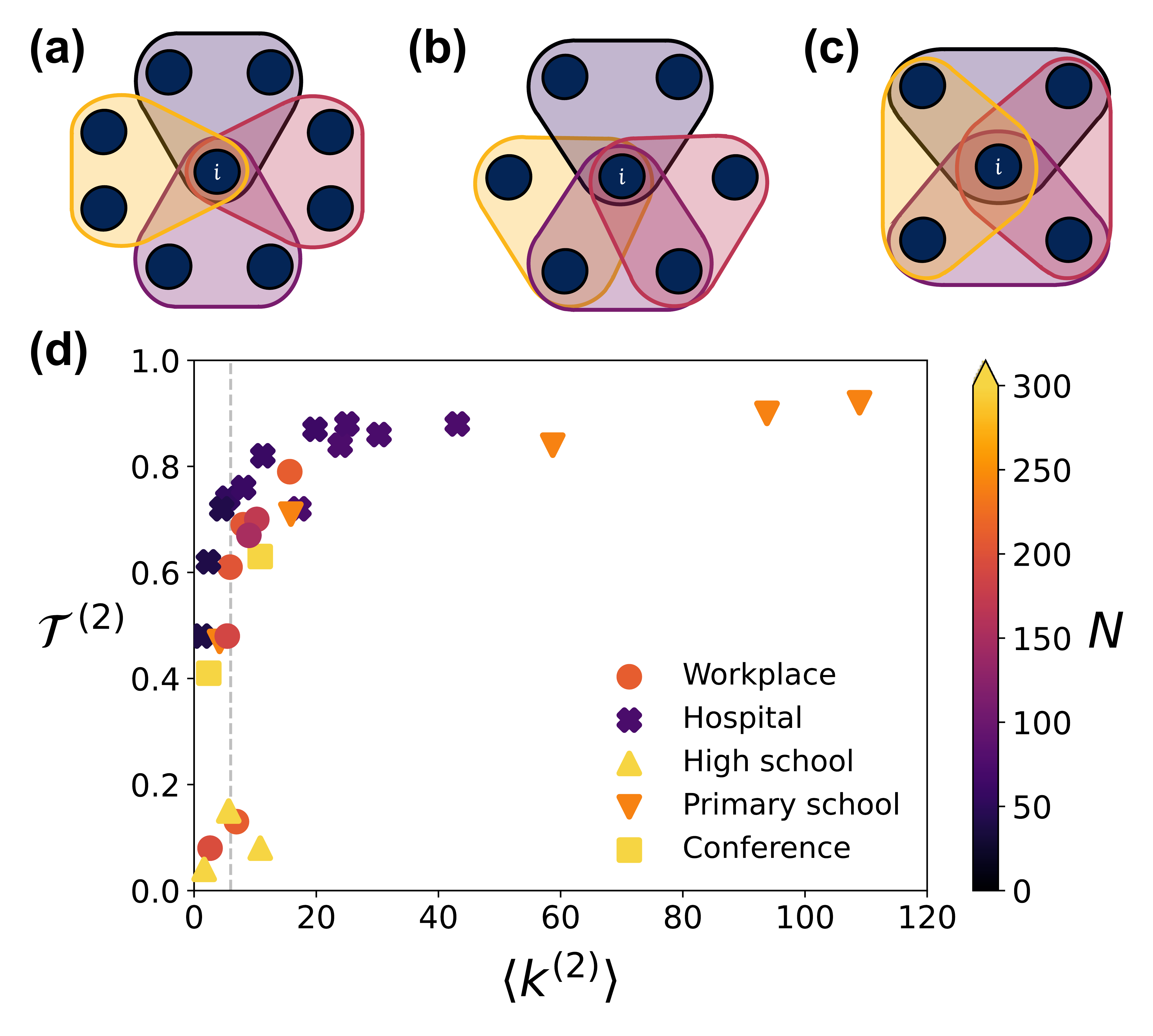}\\
\caption{{\bf Hyperedge overlap in synthetic and real-world systems}. 
(a)-(c): Examples of three configurations with different levels of node hyperedge  overlap for a node $i$ with generalized 2-degree 
$k_i^{(2)}=4$: (a) $T_i=0$; (b) $T_i=0.5$; (c) $T_i=1$. (d) Hyperedge overlap $\mathcal{T}^{(2)}$ and average generalized degree 
$\langle k^{(2)}\rangle$ of real-world hypernetworks from high-resolution face-to-face contact data. 
%
}
\label{fig1} 
\end{figure}

\medskip
\textit{Quantifying hyperedge overlap in  higher-order networks.} 
We model a system with 
higher-order interactions as 
a hypergraph $H=(\mathcal{N},\mathcal{E})$, where 
$\mathcal{N}$ is the set of $N = | \mathcal{N}|$ nodes and $\mathcal{E}$ is the set of 
hyperedges, describing the 
$E = | \mathcal{E}|$ 
interactions in groups of two or more nodes.  
Each hyperedge $e \in \mathcal{E}$ is a subset of the nodes in $\mathcal{N}$ and is characterized by its order, $m$, which is defined in terms of its cardinality, $|e|$, as $m = |e| -1$. So hyperedges of order 1 represent pairwise interactions,  interactions in  group of 3 nodes correspond to hyperdges of order 2, and so on. 
For each node $i$, $\mathcal{E}_i^{(m)}$ is the set of hyperedges $e_j$ of order $m$ such that $i\in e_j$, and $k_i^{(m)}$ is its generalized degree of order $m$, defined as the cardinality of the subset $\mathcal{E}_i^{(m)}$, i.e., $k_i^{(m)} = |\mathcal{E}_i^{(m)}|$
\cite{courtney2016generalized}.
For each order $m$, we can also define an adjacency tensor $\mathrm{A}^{(m)}$ whose generic element $a^{(m)}_{i_0,\ldots, i_m} = 1$ if the $m$-hyperedge containing nodes $i_0,\ldots, i_m$ exists, and zero otherwise. The hyperedges to which node $i$ belongs determine which nodes interact or not with $i$, and thus which nodes influence or not its dynamics. In the case of a simple graph, each link of node $i$ connects it to a distinct neighbor. Conversely, in a hypergraph, a node $i$ with two or more hyperedges of order $m\geq 2$ can share one or more neighboring nodes with different hyperedges. 
Fig.~\ref{fig1}(a)-(c) shows three different configurations for a node with four 2-hyperedges. 
The extent of overlap of the hyperedges of $i$ varies, resulting in diverse microscopic structures ranging from the non-overlapping case in panel (a) to the extreme case of maximal hyperedge overlap in (c).
To measure the local overlap among hyperedges of  order $m$ we introduce the node hyperedge overlap $T_i^{(m)}$ of a node $i$, defined for $k_i^{(m)} > 1$ as: 
\begin{equation}
    T_i^{(m)} = 1-\dfrac{ \mathcal{S}_i^{(m)} - {\mathcal{{S}}_i^{(m),-}}}{\mathcal{{S}}_i^{(m),+} -\mathcal{{S}}_i^{(m),-}},
    \label{eq:Ti}
\end{equation}
while, for $k_i^{(m)} \leq 1$, we set $T_i^{(m)}=0$ since no neighbors can overlap. In Eq.~\eqref{eq:Ti} $\mathcal{S}_i^{(m)}$ is the number of unique neighbors of node $i$ whereas $\mathcal{{S}}_i^{(m),-}$ ($\mathcal{{S}}_i^{(m),+}$) accounts for the minimum (maximum) number of unique neighbors that $i$ can have. Note that the expressions for the quantities $\mathcal{{S}}_i^{(m),+}$ and $\mathcal{{S}}_i^{(m),-}$ depend on the node generalized degree $k_i^{(m)}$, and are discussed in the SM. Similarly to the node clustering coefficient in a graph, node hyperedge overlap in Eq.~\eqref{eq:Ti} takes its minimum, i.e., $T_i^{(m)}=0$, when none of the neighbors is shared with other hyperdeges of order $m$, while it takes its maximum, i.e., $T_i^{(m)}=1$, when each neighbor is shared by two or more hyperedges of order $m$. 
For the three configurations in panels 
(a)-(c), where $m=2$, Eq.~\eqref{eq:Ti} gives $T_i^{(2)}=0$, $T_i^{(2)}=0.5$, and $T_i^{(2)}=1$, respectively. 
Finally, to characterize the degree of 
hyperedge overlap at the global scale of the whole hypergraph we average ${T}_i^{(m)}$ over all the nodes. 
\begin{equation}
    \mathcal{T}^{(m)}= \frac{\sum_ik_i^{(m)}T_i^{(m)}}{\sum_ik_i^{(m)}}.
    \label{eq:global_overlapness}
\end{equation}
\noindent 


To show that real-world systems can exhibit different levels of hyperedge overlap, we have constructed $31$ hypernetworks from $5$ high-resolution face-to-face contact data collected from various contexts~\cite{barrat_todos,SFHH,LH10,Thiers,Lyon} (more details on the datasets and how they are processed \cite{Iacopini} are described in the SM). Fig.~\ref{fig1}(d) reports the average generalized degree $\langle k^{(2)}\rangle=\sum_ik_i^{(2)}/N$ and the hyperedge overlap $\mathcal{T}^{(2)}$ for each of the hypernetworks. 
The results show that the values of $\mathcal{T}^{(2)}$ span the whole range $[0,1]$. In particular, we find that hypernetworks with the same value of $\langle k^{(2)}\rangle$ can have very different values of the hyperedge overlap $\mathcal{T}^{(2)}$. 
This is an indication of the independence of the two structural descriptors $\langle k^{(2)}\rangle$ and $\mathcal{T}^{(2)}$.
E.g., for $\langle k^{(2)}\rangle=6$ (see vertical dashed line), we observe that the hyperedge overlap $\mathcal{T}^{(2)}$ of real-world systems can vary from 0.1 to values as large as 0.8.

We will now demonstrate that this diversity of  $\mathcal{T}^{(2)}$ 
plays a crucial role in shaping the overall collective behavior of a system. 
%
%
To investigate in a systematic way the effect of the hyperedge overlap on different dynamical processes, we introduce a method to construct hypergraphs with a tunable value of $\mathcal{T}^{(2)}$. We start  from a configuration with maximum hyperedge overlap, i.e., $\mathcal{T}^{(2)}=1$, obtained by letting each node to share the same number of neighbors at each order of hyperedges (in particular, 1-hyperedges and 2-hyperedges). Then, the value of $\mathcal{T}^{(2)}$ is tuned by rewiring a fraction of the existing $2-$hyperedges, without changing the average generalized degree $\langle k^{(2)}\rangle$ (more details on the construction are provided in the SM). In this way, we obtain hypergraphs with 
fixed values of $\langle k^{(2)}\rangle$ and different degrees of hyperedge overlap, as in vertical dashed line in  
Fig.~\ref{fig1}(d). 
\begin{figure*}[t!]
\includegraphics[width=0.75\linewidth]{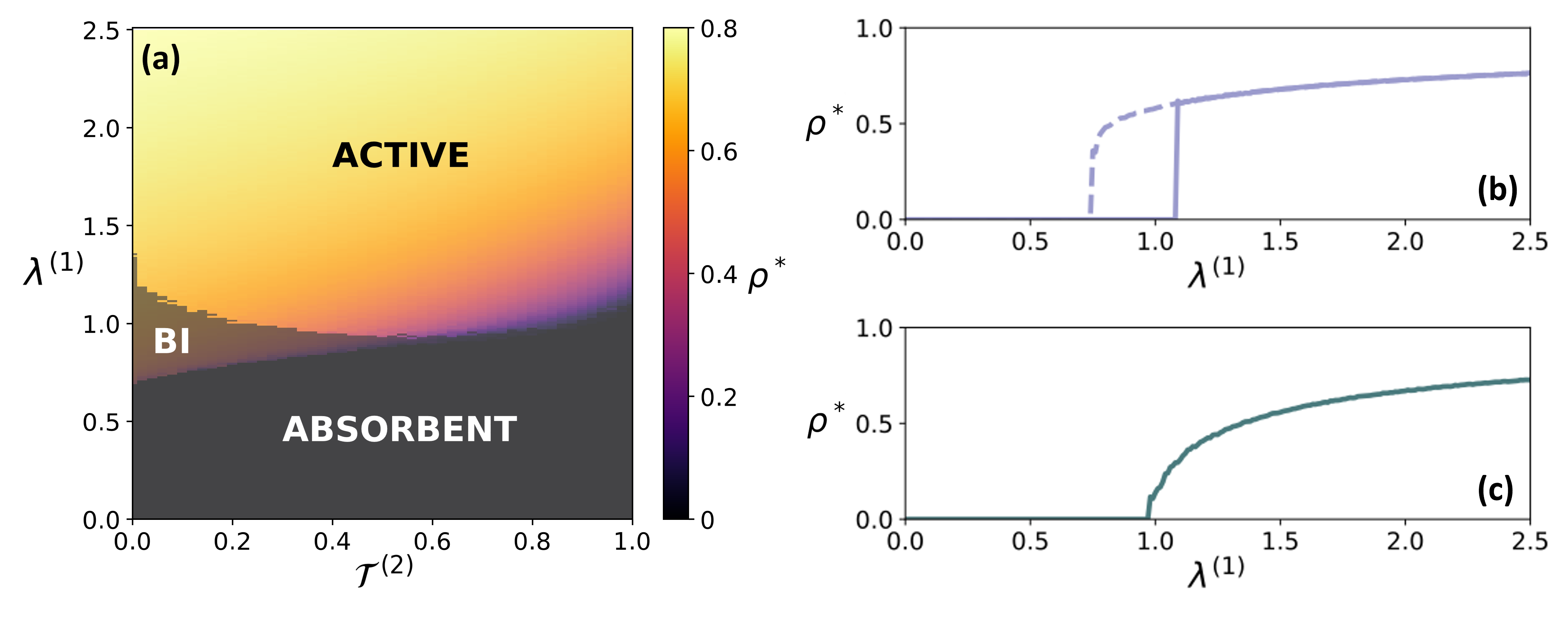}
\caption{\textbf{Effect of hyperedge overlap on complex contagion.} (a) Phase diagram for the SIS model on a 
hypergraph with $N=1000$ nodes and average generalized degrees $k^{(1)}=5$ and $k^{(2)}=6$. The value of  2-hyperedge infectivity is set to $\lambda^{(2)}=3$. 
Three phases emerge as a function 
of $\lambda^{(1)}$ and of the hyperedge  overlap $\mathcal{T}^{(2)}$: an 
absorbent phase with $\rho^{\star}=0$,   an active phase with an endemic stationary state  $\rho^{\star}
\neq 0$, and a bistability phase, where the stationary state depends on the initial conditions. (b)-(c) Two cuts of the diagram: panel (b), for $\mathcal{T}^{(2)}=0.1$, shows an explosive transition, while panel (c), for $\mathcal{T}^{(2)}=0.8$, shows a continuous transition.
}
\label{fig2} 
\end{figure*}

\bigskip
\textit{Universal effect of hyperedge overlap on dynamics.} 
We focus on two different types of dynamical processes on hypergraphs, namely complex contagion and synchronization of phase oscillators.
%
In the first case study, we consider a Susceptible-Infected-Susceptible (SIS) compartmental model for mimicking the spread of ideas through social contagion in  hypergraphs~\cite{Iacopini,de2020social}. In this framework, each node of the hypergraph represents an individual that can be either susceptible (S) or infected (I). The transition from the state S to I occurs when a susceptible individual enters in contact with infected ones, via a pairwise or a three-body interaction. 
More specifically, a susceptible individual can become infected, with probability $\beta^{(1)}$, through a 1-hyperedge (a link) with an infected individual, and with probability $\beta^{(2)}$, through a 2-hyperedge (an interaction in a group of three) with two other infected nodes. As in the standard SIS model, an  infected individual can recover with probability $\mu$. 
%
In absence of higher-order interactions, the SIS model undergoes a continuous transition from an absorbent state with a vanishing stationary fraction of infected individuals, $\rho^{\star}=0$, to an active state with $\rho^{\star}\neq0$ \cite{kiss2017mathematics}.  
In previous works~\cite{Iacopini,st2022influential}, it has been shown that the presence of higher-order interactions changes the nature of the transition from continuous to discontinuous, hence leading to the emergence of a region of bistability, where both the absorbent and active states co-exist. Here, we show that this does not occur for any hypergraph, but strongly depends on 
the level of hyperedge overlap.

Figure~\ref{fig2} shows the results of stochastic simulations of the SIS model on hypergraphs with tunable hyperedge overlap. 
Panel (a) reports the fraction $\rho^{\star}$ of infected individuals in the stationary state  as a function of $\mathcal{T}^{(2)}$ and of the rescaled transmission probability 
$\lambda^{(1)}= \langle k^{(1)}\rangle \beta^{(1)}/\mu$, while keeping the rescaled transmission probability through 2-hyperedges fixed to $\lambda^{(2)}=\langle k^{(2)} \rangle \beta^{(2)}/\mu=3$.
The phase diagram of Fig.~\ref{fig2}(a) was obtained by averaging $M=200$ simulations, half of which started with a small fraction of infected individuals $\rho(0) = 0.01$, and the other half with $\rho(0) = 0.8$.
We observe the presence of three phases, with a region of bistability that only appears for small values of $\mathcal{T}^{(2)}$. An example of the explosive transition associated to bistability is shown in Fig.~\ref{fig2}(b), which illustrates  $\rho^{\star}$ vs. $\lambda^{(1)}$ for $\mathcal{T}^{(2)}=0.1$. 
The three phases merge at a tricritical point, leading to the disappearance of the bistability. Beyond this point, the transition from the absorbent to the active phase is continuous. As an example, Fig.~\ref{fig2}(c) illustrates the continuous behavior of $\rho^{\star}$ vs. $\lambda^{(1)}$ in a hypergraph with $\mathcal{T}^{(2)}=0.8$. 
In this case, the high overlap of the 
hyperedges of a node reduces the effectiveness in propagating its infected state.

\begin{figure*}[t!]
\includegraphics[width=0.75\linewidth]{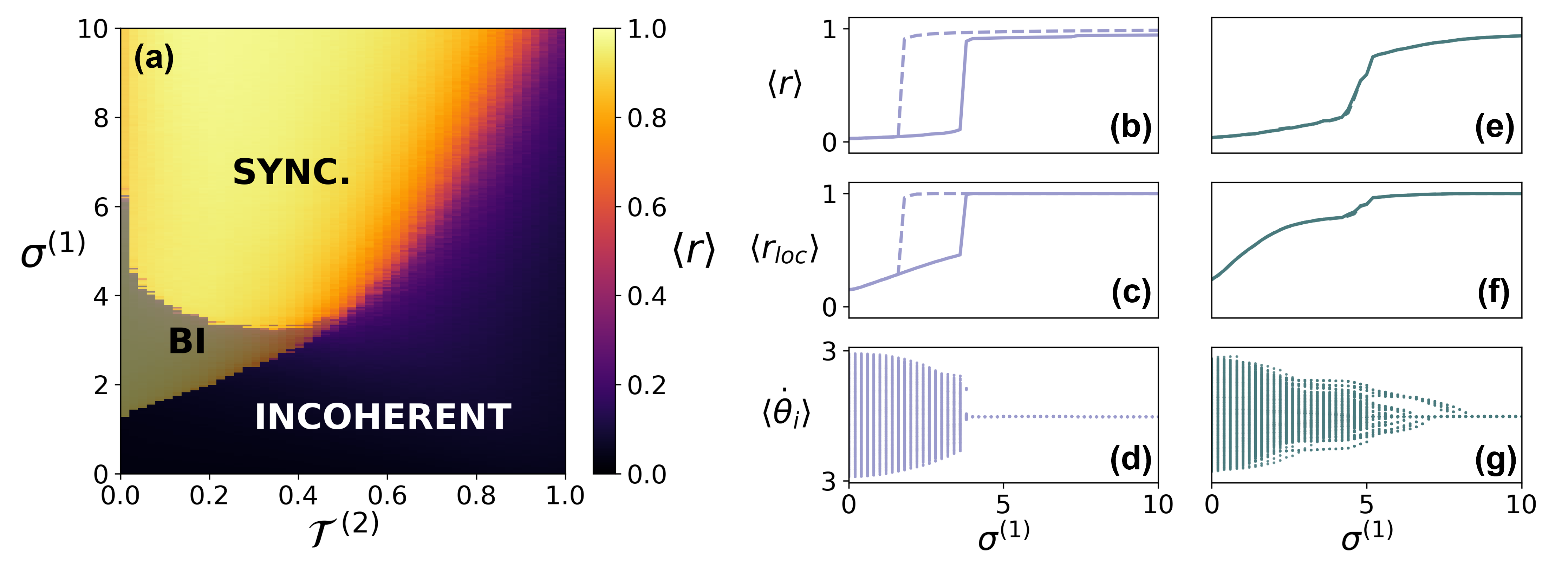}\\
\caption{\textbf{Effect of hyperedge overlap on synchronization.} (a) Phase diagram for the Kuramoto model on a hypergraph with $N=1000$ nodes and average generalized degrees $k^{(1)}=5$ and $k^{(2)}=6$. The coupling strength for three-body interactions is set to $\sigma^{(2)}=3$. 
Three phases emerge as a function of $\sigma^{(1)}$ and 
$\mathcal{T}^{(2)}$: an incoherent phase with low values of $\langle r \rangle$, a synchronized phase with large 
$\langle r \rangle$, and a bistability phase where the system can be synchronized or not depending on the initial conditions. 
(b)-(g) Two cuts of the diagram: panels (b)-(d), for $\mathcal{T}^{(2)}=0.06$, show an explosive transition, while panels (e)-(g), for $\mathcal{T}^{(2)}=0.6$, show a continuous transition. 
Panels (b) and (d) illustrate the order parameter $\langle r \rangle$, while panels (c) and (f) the local synchronization parameter $\langle r_{loc} \rangle$. Panels (d) and (f) show the average effective frequencies $\langle \dot\theta_i \rangle $ of the hypergraph nodes. 
}
\label{fig3} 
\end{figure*}

%
As a second case study, we consider synchronization dynamics. In more details, we investigate synchronization in a system of Kuramoto oscillators \cite{kuramoto1975self} coupled via pairwise and three-body interactions. Each node $i$ ($i=1,\ldots,N$) of the system is a phase oscillator and is characterized, at time $t$, by its phase $\theta_i(t)$.  
The time evolution of the system is governed by the set of coupled equations 
\cite{arenas_synch,higher-order_kuramoto}:
\begin{multline}
\dot\theta_i=\omega_i+
\frac{\sigma^{(1)}}{\langle k^{(1)}\rangle}\sum_{j=1}^{N}a^{(1)}_{ij}\sin(\theta_j-\theta_i)+\\
+\frac{\sigma^{(2)}}{2!\langle k^{(2)}\rangle}\sum_{j=1}^{N}\sum_{k=1}^{N}a^{(2)}_{ijk}\sin(\theta_j+\theta_k-2\theta_i),
\label{eq:kuramoto}
\end{multline}
where $\omega_i$ is the natural frequency associated to oscillator $i$, which is randomly sampled from a uniform distribution.  
The quantities $\sigma^{(1)}$ and $\sigma^{(2)}$ represent the coupling strength for pairwise (order 1) and three-body (order 2) interactions, respectively, while $a^{(1)}_{ij}=1$
if there is a link between node $i$ and $j$, and $a^{(2)}_{ijk}=1$ if the hypergraph has a hyperedge containing $i$, $j$ and $k$. 
The synchronization transition is captured by the order parameter $r(t)={N}^{-1}\left|\sum_{j=1}^Ne^{\mathrm{i}\theta_j(t)}\right|$, ranging from 0 (incoherence) to 1 (full synchronization). With only pairwise interactions, and in the absence of additional  ingredients~\cite{explosive_raissa}, a continuous synchronization transition takes place when the coupling strength $\sigma^{(1)}$ exceeds a critical value, $\sigma^{(1)}_c$ \cite{acebron2005kuramoto}. 
As in the case of the SIS dynamics we have studied this model on hypergraphs with tunable hyperedge overlap.
Figure~\ref{fig3} shows the results of the numerical integration of Eqs.~\eqref{eq:kuramoto} for $\sigma^{(2)}=3$ and for various values of $\sigma^{(1)}$ and $\mathcal{T}^{(2)}$. 
For each set of parameters, we perform $M=200$ simulations, half with a narrow ($\Delta\theta = 0.02$) and half with a wider  ($\Delta\theta = 2\pi$) distribution of initial phases.  
For each simulation, we let the system evolve a transient time $t_r$ and calculate $\langle r\rangle=\frac{1}{\Delta t}\int_{t_r}^{t_r+\Delta t}r(t)dt$, where $\Delta t\rightarrow\infty$.
We then consider the median of $\langle r\rangle$
over all runs.
The phase diagram in Fig.~\ref{fig3}(a) reporting $\langle r\rangle$ as a function of $\mathcal{T}^{(2)}$ and 
$\sigma^{(2)}$ resembles that of the SIS model in Fig.~\ref{fig2}(a). Again the region of bistability is present for small $\mathcal{T}^{(2)}$, but disappears for large values. 
Fig.~\ref{fig3}(b) and (e) show $\langle r\rangle$ as a function of $\sigma^{(1)}$, respectively  
for $\mathcal{T}^{(2)}=0.06$, where the transition to synchronization is explosive, and for $\mathcal{T}^{(2)}=0.6$, where the  transition is continuous. 
To gain a better insight into the microscopic mechanisms underlying the two different transitions to synchronization of Eq.~\eqref{eq:kuramoto}, 
we introduce the following measure of higher-order local synchronization:
\begin{multline}
\nonumber
\langle r_{loc} \rangle=\frac{1}{\sum_{i=1}\left(\sum_{j=1}a^{(1)}_{ij}+\frac{1}{2!}\sum_{j=1}\sum_{k=1}a^{(2)}_{ijk}\right)}\times\\
\sum_{i=1}\Bigg(\sum_{j=1}\left\vert\lim_{\Delta t\rightarrow\infty}\frac{a^{(1)}_{ij}}{\Delta t}\int_{t_r}^{t_r+\Delta t}e^{\mathrm{i}[\theta_j(t)-\theta_i(t)]}dt\right\vert+\\
\frac{1}{2!}\sum_{j=1}\sum_{k=1}\left\vert\lim_{\Delta t\rightarrow\infty}\frac{a^{(2)}_{ijk}}{\Delta t}\int_{t_r}^{t_r+\Delta t}e^{\mathrm{i}[\theta_j(t)+\theta_k(t)-2\theta_i(t)]}dt\right\vert \vphantom{\frac12}\Bigg).
\label{eq:rlocal}
\end{multline}
This definition of $\langle r_{loc} \rangle$ extends the one used for network edges \cite{rlink}, since it gathers the phase coherence between pairs of nodes connected by links, as well as among triplets of nodes connected by 2-hyperedges.
As for the order parameter $\langle r \rangle$, full synchronization corresponds to $\langle r_{loc}\rangle=1$, while complete incoherence yields $\langle r_{loc} \rangle=0$. Fig.~\ref{fig3}(c) and (f) 
show that the hyperedge overlap radically changes the behavior of $\langle r_{loc} \rangle$ as a function of $\sigma^{(1)}$. 
In agreement with $\langle r \rangle$, also $\langle r_{loc} \rangle$ indicates that the transition is explosive for  hypergraphs with low hyperedge overlap, while it is continuous for hypergraphs with high hyperedge overlap.  
However, at variance with $\langle r\rangle$, for large values of $\sigma^{(1)}$, $\langle r_{loc} \rangle$ 
reaches the value $\langle r_{loc} \rangle=1$, meaning that, even if the system is not globally synchronized, there is still a high degree of local synchronization. The complete phase diagram for $\langle r_{loc} \rangle$ as function of $\sigma^{(1)}$ and $\mathcal{T}^{(2)}$ is reported in the SM.
The high degree of local synchronization is a consequence of the formation of synchronous clusters in the system, and depends on the level of hyperedge overlap.   
This effect becomes evident from the analysis of the average effective frequencies of the oscillators, defined as: 
$ \langle \dot\theta_i \rangle 
=\frac{1}{\Delta t}\int_{t_r}^{t_r+\Delta t}\dot\theta_i(t)dt$, where $\Delta t\rightarrow\infty$.
In Fig.~\ref{fig3}(d) and (g) we 
compare the effective frequencies $\langle \dot\theta_i \rangle$ in hypergraphs with different levels of hyperedge overlap. 
When $\mathcal{T}^{(2)}$ is small, as in Fig.~\ref{fig3}(d),
the 2-hyperedges have few nodes in commmon  and this makes difficult the nucleation of synchronous clusters. Consequently, the synchronization onset is abrupt and, as we increase $\sigma^{(1)}$, the effective frequencies 
$\langle \dot\theta_i \rangle$  
converge suddenly to their mean value. 
Conversely, when $\mathcal{T}^{(2)}$ is small, as in Fig.~\ref{fig3}(g), the 2-hyperedges share multiple nodes, which promotes the formation of synchronous clusters. The result is that, as $\sigma^{(1)}$ increases from zero, we find several groups of nodes locked at the same effective frequency. As $\sigma^{(1)}$ further increases, these groups smoothly merge together, until the whole system becomes synchronized.

\textit{Conclusions.}  In this Letter, we have analyzed the effects of the structural organization of higher-order interactions in hypergraphs. To this aim we have introduced a novel metric for higher-order networks called hyperedge overlap, quantifying the degree of shared nodes among hyperedges. First, we found a diverse range of hyperedge overlap values in real hypergraphs, highlighting the structural variations within these systems. 
Then, we examined the effects of hyperedge overlap in shaping the onset of collective phenomena, discovering that it plays a universal role in hindering the explosivity and bistability often associated to higher-order structures. 
Thus, we have shown that the presence of higher-order interactions alone does not guarantee the existence of an abrupt transition, as it strongly depends on how hyperedges are organized at the microscopic level. 





\paragraph*{Acknowledgements.} S.L.O and J.G.G. acknowledge financial support from the Departamento de Industria e Innovaci\'on del Gobierno de Arag\'on y Fondo Social Europeo (FENOL group grant E36-23R) and from Ministerio de Ciencia e Innovaci\'on (grant PID2020-113582GB-I00). V.L. acknowledges support from the PNRR GRInS Project. 


\begin{thebibliography}{35}
\expandafter\ifx\csname natexlab\endcsname\relax\def\natexlab#1{#1}\fi
\expandafter\ifx\csname bibnamefont\endcsname\relax
  \def\bibnamefont#1{#1}\fi
\expandafter\ifx\csname bibfnamefont\endcsname\relax
  \def\bibfnamefont#1{#1}\fi
\expandafter\ifx\csname citenamefont\endcsname\relax
  \def\citenamefont#1{#1}\fi
\expandafter\ifx\csname url\endcsname\relax
  \def\url#1{\texttt{#1}}\fi
\expandafter\ifx\csname urlprefix\endcsname\relax\def\urlprefix{URL }\fi
\providecommand{\bibinfo}[2]{#2}
\providecommand{\eprint}[2][]{\url{#2}}

\bibitem[{\citenamefont{Newman}(2003)}]{newman}
\bibinfo{author}{\bibfnamefont{M.~E.~J.} \bibnamefont{Newman}},
  \bibinfo{journal}{SIAM Review} \textbf{\bibinfo{volume}{45}},
  \bibinfo{pages}{167} (\bibinfo{year}{2003}).

\bibitem[{\citenamefont{Latora et~al.}(2017)\citenamefont{Latora, Nicosia, and
  Russo}}]{latorabook}
\bibinfo{author}{\bibfnamefont{V.}~\bibnamefont{Latora}},
  \bibinfo{author}{\bibfnamefont{V.}~\bibnamefont{Nicosia}}, \bibnamefont{and}
  \bibinfo{author}{\bibfnamefont{G.}~\bibnamefont{Russo}},
  \emph{\bibinfo{title}{Complex Networks: Principles, Methods and
  Applications}} (\bibinfo{publisher}{Cambridge University Press},
  \bibinfo{address}{USA}, \bibinfo{year}{2017}), \bibinfo{edition}{1st} ed.,
  ISBN \bibinfo{isbn}{1107103185}.

\bibitem[{\citenamefont{Barrat et~al.}(2008)\citenamefont{Barrat, Barthlemy,
  and Vespignani}}]{barrat_book_2008}
\bibinfo{author}{\bibfnamefont{A.}~\bibnamefont{Barrat}},
  \bibinfo{author}{\bibfnamefont{M.}~\bibnamefont{Barthlemy}},
  \bibnamefont{and}
  \bibinfo{author}{\bibfnamefont{A.}~\bibnamefont{Vespignani}},
  \emph{\bibinfo{title}{Dynamical Processes on Complex Networks}}
  (\bibinfo{publisher}{Cambridge University Press}, \bibinfo{address}{New York,
  NY, USA}, \bibinfo{year}{2008}), ISBN \bibinfo{isbn}{0521879507,
  9780521879507}.

\bibitem[{\citenamefont{Christensen and Moloney}(2005)}]{Percolation}
\bibinfo{author}{\bibfnamefont{K.}~\bibnamefont{Christensen}} \bibnamefont{and}
  \bibinfo{author}{\bibfnamefont{N.~R.} \bibnamefont{Moloney}}, in
  \emph{\bibinfo{booktitle}{Complexity and Criticality}}
  (\bibinfo{year}{2005}).

\bibitem[{\citenamefont{Arenas et~al.}(2008)\citenamefont{Arenas,
  Díaz-Guilera, Kurths, Moreno, and Zhou}}]{ARENAS200893}
\bibinfo{author}{\bibfnamefont{A.}~\bibnamefont{Arenas}},
  \bibinfo{author}{\bibfnamefont{A.}~\bibnamefont{Díaz-Guilera}},
  \bibinfo{author}{\bibfnamefont{J.}~\bibnamefont{Kurths}},
  \bibinfo{author}{\bibfnamefont{Y.}~\bibnamefont{Moreno}}, \bibnamefont{and}
  \bibinfo{author}{\bibfnamefont{C.}~\bibnamefont{Zhou}},
  \bibinfo{journal}{Physics Reports} \textbf{\bibinfo{volume}{469}},
  \bibinfo{pages}{93} (\bibinfo{year}{2008}), ISSN \bibinfo{issn}{0370-1573}.

\bibitem[{\citenamefont{Pastor-Satorras
  et~al.}(2015)\citenamefont{Pastor-Satorras, Castellano, Van~Mieghem, and
  Vespignani}}]{RevModPhys.87.925}
\bibinfo{author}{\bibfnamefont{R.}~\bibnamefont{Pastor-Satorras}},
  \bibinfo{author}{\bibfnamefont{C.}~\bibnamefont{Castellano}},
  \bibinfo{author}{\bibfnamefont{P.}~\bibnamefont{Van~Mieghem}},
  \bibnamefont{and}
  \bibinfo{author}{\bibfnamefont{A.}~\bibnamefont{Vespignani}},
  \bibinfo{journal}{Rev. Mod. Phys.} \textbf{\bibinfo{volume}{87}},
  \bibinfo{pages}{925} (\bibinfo{year}{2015}).

\bibitem[{\citenamefont{Szabó and Fáth}(2007)}]{SZABO200797}
\bibinfo{author}{\bibfnamefont{G.}~\bibnamefont{Szabó}} \bibnamefont{and}
  \bibinfo{author}{\bibfnamefont{G.}~\bibnamefont{Fáth}},
  \bibinfo{journal}{Physics Reports} \textbf{\bibinfo{volume}{446}},
  \bibinfo{pages}{97} (\bibinfo{year}{2007}), ISSN \bibinfo{issn}{0370-1573}.

\bibitem[{\citenamefont{Battiston et~al.}(2020)\citenamefont{Battiston,
  Cencetti, Iacopini, Latora, Lucas, Patania, Young, and Petri}}]{report_Ho}
\bibinfo{author}{\bibfnamefont{F.}~\bibnamefont{Battiston}},
  \bibinfo{author}{\bibfnamefont{G.}~\bibnamefont{Cencetti}},
  \bibinfo{author}{\bibfnamefont{I.}~\bibnamefont{Iacopini}},
  \bibinfo{author}{\bibfnamefont{V.}~\bibnamefont{Latora}},
  \bibinfo{author}{\bibfnamefont{M.}~\bibnamefont{Lucas}},
  \bibinfo{author}{\bibfnamefont{A.}~\bibnamefont{Patania}},
  \bibinfo{author}{\bibfnamefont{J.-G.} \bibnamefont{Young}}, \bibnamefont{and}
  \bibinfo{author}{\bibfnamefont{G.}~\bibnamefont{Petri}},
  \bibinfo{journal}{Physics Reports} \textbf{\bibinfo{volume}{874}},
  \bibinfo{pages}{1} (\bibinfo{year}{2020}), ISSN \bibinfo{issn}{0370-1573},
  \bibinfo{note}{networks beyond pairwise interactions: Structure and
  dynamics}.

\bibitem[{\citenamefont{Majhi~Soumen and Dibakar}(2022)}]{otra_review}
\bibinfo{author}{\bibfnamefont{P.~M.} \bibnamefont{Majhi~Soumen}}
  \bibnamefont{and} \bibinfo{author}{\bibfnamefont{G.}~\bibnamefont{Dibakar}},
  \bibinfo{journal}{J. R. Soc. Interface.} \textbf{\bibinfo{volume}{19}}
  (\bibinfo{year}{2022}).

\bibitem[{\citenamefont{Bick et~al.}(2022)\citenamefont{Bick, Gross,
  Harrington, and Schaub}}]{bick2022higherorder}
\bibinfo{author}{\bibfnamefont{C.}~\bibnamefont{Bick}},
  \bibinfo{author}{\bibfnamefont{E.}~\bibnamefont{Gross}},
  \bibinfo{author}{\bibfnamefont{H.~A.} \bibnamefont{Harrington}},
  \bibnamefont{and} \bibinfo{author}{\bibfnamefont{M.~T.} \bibnamefont{Schaub}}
  (\bibinfo{year}{2022}), \eprint{2104.11329}.

\bibitem[{\citenamefont{Iacopini et~al.}(2019)\citenamefont{Iacopini, Petri,
  Barrat, and Latora}}]{Iacopini}
\bibinfo{author}{\bibfnamefont{I.}~\bibnamefont{Iacopini}},
  \bibinfo{author}{\bibfnamefont{G.}~\bibnamefont{Petri}},
  \bibinfo{author}{\bibfnamefont{A.}~\bibnamefont{Barrat}}, \bibnamefont{and}
  \bibinfo{author}{\bibfnamefont{V.}~\bibnamefont{Latora}},
  \bibinfo{journal}{Nat Commun} \textbf{\bibinfo{volume}{10}},
  \bibinfo{pages}{2485} (\bibinfo{year}{2019}).

\bibitem[{\citenamefont{Matamalas et~al.}(2020)\citenamefont{Matamalas,
  G\'omez, and Arenas}}]{arenas_expl}
\bibinfo{author}{\bibfnamefont{J.~T.} \bibnamefont{Matamalas}},
  \bibinfo{author}{\bibfnamefont{S.}~\bibnamefont{G\'omez}}, \bibnamefont{and}
  \bibinfo{author}{\bibfnamefont{A.}~\bibnamefont{Arenas}},
  \bibinfo{journal}{Phys. Rev. Res.} \textbf{\bibinfo{volume}{2}},
  \bibinfo{pages}{012049} (\bibinfo{year}{2020}).

\bibitem[{\citenamefont{de~Arruda
  et~al.}(2020{\natexlab{a}})\citenamefont{de~Arruda, Petri, and
  Moreno}}]{yamir_hg}
\bibinfo{author}{\bibfnamefont{G.~F.} \bibnamefont{de~Arruda}},
  \bibinfo{author}{\bibfnamefont{G.}~\bibnamefont{Petri}}, \bibnamefont{and}
  \bibinfo{author}{\bibfnamefont{Y.}~\bibnamefont{Moreno}},
  \bibinfo{journal}{Phys. Rev. Res.} \textbf{\bibinfo{volume}{2}},
  \bibinfo{pages}{023032} (\bibinfo{year}{2020}{\natexlab{a}}).

\bibitem[{\citenamefont{Tanaka and Aoyagi}(2011)}]{hosynch}
\bibinfo{author}{\bibfnamefont{T.}~\bibnamefont{Tanaka}} \bibnamefont{and}
  \bibinfo{author}{\bibfnamefont{T.}~\bibnamefont{Aoyagi}},
  \bibinfo{journal}{Phys. Rev. Lett.} \textbf{\bibinfo{volume}{106}},
  \bibinfo{pages}{224101} (\bibinfo{year}{2011}).

\bibitem[{\citenamefont{Mill\'an et~al.}(2020)\citenamefont{Mill\'an, Torres,
  and Bianconi}}]{exp2}
\bibinfo{author}{\bibfnamefont{A.~P.} \bibnamefont{Mill\'an}},
  \bibinfo{author}{\bibfnamefont{J.~J.} \bibnamefont{Torres}},
  \bibnamefont{and} \bibinfo{author}{\bibfnamefont{G.}~\bibnamefont{Bianconi}},
  \bibinfo{journal}{Phys. Rev. Lett.} \textbf{\bibinfo{volume}{124}},
  \bibinfo{pages}{218301} (\bibinfo{year}{2020}).

\bibitem[{\citenamefont{Skardal and Arenas}(2020)}]{arenas_synch}
\bibinfo{author}{\bibfnamefont{P.}~\bibnamefont{Skardal}} \bibnamefont{and}
  \bibinfo{author}{\bibfnamefont{A.}~\bibnamefont{Arenas}},
  \bibinfo{journal}{Commun Phys} \textbf{\bibinfo{volume}{3}}
  (\bibinfo{year}{2020}).

\bibitem[{\citenamefont{Gambuzza et~al.}(2021)\citenamefont{Gambuzza, Di~Patti,
  Gallo, Lepri, Romance, Criado, Frasca, Latora, and
  Boccaletti}}]{gambuzza2021stability}
\bibinfo{author}{\bibfnamefont{L.~V.} \bibnamefont{Gambuzza}},
  \bibinfo{author}{\bibfnamefont{F.}~\bibnamefont{Di~Patti}},
  \bibinfo{author}{\bibfnamefont{L.}~\bibnamefont{Gallo}},
  \bibinfo{author}{\bibfnamefont{S.}~\bibnamefont{Lepri}},
  \bibinfo{author}{\bibfnamefont{M.}~\bibnamefont{Romance}},
  \bibinfo{author}{\bibfnamefont{R.}~\bibnamefont{Criado}},
  \bibinfo{author}{\bibfnamefont{M.}~\bibnamefont{Frasca}},
  \bibinfo{author}{\bibfnamefont{V.}~\bibnamefont{Latora}}, \bibnamefont{and}
  \bibinfo{author}{\bibfnamefont{S.}~\bibnamefont{Boccaletti}},
  \bibinfo{journal}{Nature communications} \textbf{\bibinfo{volume}{12}},
  \bibinfo{pages}{1255} (\bibinfo{year}{2021}).

\bibitem[{\citenamefont{Gallo et~al.}(2022)\citenamefont{Gallo, Muolo,
  Gambuzza, Latora, Frasca, and Carletti}}]{gallo2022synchronization}
\bibinfo{author}{\bibfnamefont{L.}~\bibnamefont{Gallo}},
  \bibinfo{author}{\bibfnamefont{R.}~\bibnamefont{Muolo}},
  \bibinfo{author}{\bibfnamefont{L.~V.} \bibnamefont{Gambuzza}},
  \bibinfo{author}{\bibfnamefont{V.}~\bibnamefont{Latora}},
  \bibinfo{author}{\bibfnamefont{M.}~\bibnamefont{Frasca}}, \bibnamefont{and}
  \bibinfo{author}{\bibfnamefont{T.}~\bibnamefont{Carletti}},
  \bibinfo{journal}{Communications Physics} \textbf{\bibinfo{volume}{5}},
  \bibinfo{pages}{263} (\bibinfo{year}{2022}).

\bibitem[{\citenamefont{Civilini et~al.}(2023)\citenamefont{Civilini, Sadekar,
  Battiston, Gómez-Gardeñes, and Latora}}]{civilini2023explosive}
\bibinfo{author}{\bibfnamefont{A.}~\bibnamefont{Civilini}},
  \bibinfo{author}{\bibfnamefont{O.}~\bibnamefont{Sadekar}},
  \bibinfo{author}{\bibfnamefont{F.}~\bibnamefont{Battiston}},
  \bibinfo{author}{\bibfnamefont{J.}~\bibnamefont{Gómez-Gardeñes}},
  \bibnamefont{and} \bibinfo{author}{\bibfnamefont{V.}~\bibnamefont{Latora}}
  (\bibinfo{year}{2023}), \eprint{2303.11475}.

\bibitem[{\citenamefont{Battiston et~al.}(2021)\citenamefont{Battiston, Amico,
  Barrat, Bianconi, Ferraz~de Arruda, Franceschiello, Iacopini, K{\'e}fi,
  Latora, Moreno et~al.}}]{nature}
\bibinfo{author}{\bibfnamefont{F.}~\bibnamefont{Battiston}},
  \bibinfo{author}{\bibfnamefont{E.}~\bibnamefont{Amico}},
  \bibinfo{author}{\bibfnamefont{A.}~\bibnamefont{Barrat}},
  \bibinfo{author}{\bibfnamefont{G.}~\bibnamefont{Bianconi}},
  \bibinfo{author}{\bibfnamefont{G.}~\bibnamefont{Ferraz~de Arruda}},
  \bibinfo{author}{\bibfnamefont{B.}~\bibnamefont{Franceschiello}},
  \bibinfo{author}{\bibfnamefont{I.}~\bibnamefont{Iacopini}},
  \bibinfo{author}{\bibfnamefont{S.}~\bibnamefont{K{\'e}fi}},
  \bibinfo{author}{\bibfnamefont{V.}~\bibnamefont{Latora}},
  \bibinfo{author}{\bibfnamefont{Y.}~\bibnamefont{Moreno}},
  \bibnamefont{et~al.}, \bibinfo{journal}{Nature Physics}
  \textbf{\bibinfo{volume}{17}}, \bibinfo{pages}{1093} (\bibinfo{year}{2021}).

\bibitem[{\citenamefont{Kuehn and Bick}(2021)}]{bick}
\bibinfo{author}{\bibfnamefont{C.}~\bibnamefont{Kuehn}} \bibnamefont{and}
  \bibinfo{author}{\bibfnamefont{C.}~\bibnamefont{Bick}},
  \bibinfo{journal}{Science Advances} \textbf{\bibinfo{volume}{7}},
  \bibinfo{pages}{eabe3824} (\bibinfo{year}{2021}).

\bibitem[{\citenamefont{Courtney and Bianconi}(2016)}]{courtney2016generalized}
\bibinfo{author}{\bibfnamefont{O.~T.} \bibnamefont{Courtney}} \bibnamefont{and}
  \bibinfo{author}{\bibfnamefont{G.}~\bibnamefont{Bianconi}},
  \bibinfo{journal}{Physical Review E} \textbf{\bibinfo{volume}{93}},
  \bibinfo{pages}{062311} (\bibinfo{year}{2016}).

\bibitem[{\citenamefont{Génois and Barrat}(2018)}]{barrat_todos}
\bibinfo{author}{\bibfnamefont{M.}~\bibnamefont{Génois}} \bibnamefont{and}
  \bibinfo{author}{\bibfnamefont{A.}~\bibnamefont{Barrat}},
  \bibinfo{journal}{EPJ Data Science} \textbf{\bibinfo{volume}{7}},
  \bibinfo{pages}{1} (\bibinfo{year}{2018}).

\bibitem[{\citenamefont{Isella et~al.}(2011)\citenamefont{Isella, Stehlé,
  Barrat, Cattuto, Pinton, and Van~den Broeck}}]{SFHH}
\bibinfo{author}{\bibfnamefont{L.}~\bibnamefont{Isella}},
  \bibinfo{author}{\bibfnamefont{J.}~\bibnamefont{Stehlé}},
  \bibinfo{author}{\bibfnamefont{A.}~\bibnamefont{Barrat}},
  \bibinfo{author}{\bibfnamefont{C.}~\bibnamefont{Cattuto}},
  \bibinfo{author}{\bibfnamefont{J.-F.} \bibnamefont{Pinton}},
  \bibnamefont{and} \bibinfo{author}{\bibfnamefont{W.}~\bibnamefont{Van~den
  Broeck}}, \bibinfo{journal}{Journal of Theoretical Biology}
  \textbf{\bibinfo{volume}{271}}, \bibinfo{pages}{166} (\bibinfo{year}{2011}).

\bibitem[{\citenamefont{Vanhems et~al.}(2013)\citenamefont{Vanhems, Barrat,
  Cattuto, Pinton, Khanafer, Régis, Kim, Comte, and Voirin}}]{LH10}
\bibinfo{author}{\bibfnamefont{P.}~\bibnamefont{Vanhems}},
  \bibinfo{author}{\bibfnamefont{A.}~\bibnamefont{Barrat}},
  \bibinfo{author}{\bibfnamefont{C.}~\bibnamefont{Cattuto}},
  \bibinfo{author}{\bibfnamefont{J.-F.} \bibnamefont{Pinton}},
  \bibinfo{author}{\bibfnamefont{N.}~\bibnamefont{Khanafer}},
  \bibinfo{author}{\bibfnamefont{C.}~\bibnamefont{Régis}},
  \bibinfo{author}{\bibfnamefont{B.-a.} \bibnamefont{Kim}},
  \bibinfo{author}{\bibfnamefont{B.}~\bibnamefont{Comte}}, \bibnamefont{and}
  \bibinfo{author}{\bibfnamefont{N.}~\bibnamefont{Voirin}},
  \bibinfo{journal}{PLOS ONE} \textbf{\bibinfo{volume}{8}}, \bibinfo{pages}{1}
  (\bibinfo{year}{2013}).

\bibitem[{\citenamefont{Mastrandrea et~al.}(2015)\citenamefont{Mastrandrea,
  Fournet, and Barrat}}]{Thiers}
\bibinfo{author}{\bibfnamefont{R.}~\bibnamefont{Mastrandrea}},
  \bibinfo{author}{\bibfnamefont{J.}~\bibnamefont{Fournet}}, \bibnamefont{and}
  \bibinfo{author}{\bibfnamefont{A.}~\bibnamefont{Barrat}},
  \bibinfo{journal}{PLOS ONE} \textbf{\bibinfo{volume}{10}}, \bibinfo{pages}{1}
  (\bibinfo{year}{2015}).

\bibitem[{\citenamefont{Stehlé et~al.}(2011)\citenamefont{Stehlé, Voirin,
  Barrat, Cattuto, Isella, Pinton, Quaggiotto, Van~den Broeck, Régis, Lina
  et~al.}}]{Lyon}
\bibinfo{author}{\bibfnamefont{J.}~\bibnamefont{Stehlé}},
  \bibinfo{author}{\bibfnamefont{N.}~\bibnamefont{Voirin}},
  \bibinfo{author}{\bibfnamefont{A.}~\bibnamefont{Barrat}},
  \bibinfo{author}{\bibfnamefont{C.}~\bibnamefont{Cattuto}},
  \bibinfo{author}{\bibfnamefont{L.}~\bibnamefont{Isella}},
  \bibinfo{author}{\bibfnamefont{J.-F.} \bibnamefont{Pinton}},
  \bibinfo{author}{\bibfnamefont{M.}~\bibnamefont{Quaggiotto}},
  \bibinfo{author}{\bibfnamefont{W.}~\bibnamefont{Van~den Broeck}},
  \bibinfo{author}{\bibfnamefont{C.}~\bibnamefont{Régis}},
  \bibinfo{author}{\bibfnamefont{B.}~\bibnamefont{Lina}}, \bibnamefont{et~al.},
  \bibinfo{journal}{PLOS ONE} \textbf{\bibinfo{volume}{6}}, \bibinfo{pages}{1}
  (\bibinfo{year}{2011}).

\bibitem[{\citenamefont{de~Arruda
  et~al.}(2020{\natexlab{b}})\citenamefont{de~Arruda, Petri, and
  Moreno}}]{de2020social}
\bibinfo{author}{\bibfnamefont{G.~F.} \bibnamefont{de~Arruda}},
  \bibinfo{author}{\bibfnamefont{G.}~\bibnamefont{Petri}}, \bibnamefont{and}
  \bibinfo{author}{\bibfnamefont{Y.}~\bibnamefont{Moreno}},
  \bibinfo{journal}{Physical Review Research} \textbf{\bibinfo{volume}{2}},
  \bibinfo{pages}{023032} (\bibinfo{year}{2020}{\natexlab{b}}).

\bibitem[{\citenamefont{Kiss et~al.}(2017)\citenamefont{Kiss, Miller, Simon
  et~al.}}]{kiss2017mathematics}
\bibinfo{author}{\bibfnamefont{I.~Z.} \bibnamefont{Kiss}},
  \bibinfo{author}{\bibfnamefont{J.~C.} \bibnamefont{Miller}},
  \bibinfo{author}{\bibfnamefont{P.~L.} \bibnamefont{Simon}},
  \bibnamefont{et~al.}, \bibinfo{journal}{Cham: Springer}
  \textbf{\bibinfo{volume}{598}}, \bibinfo{pages}{31} (\bibinfo{year}{2017}).

\bibitem[{\citenamefont{St-Onge et~al.}(2022)\citenamefont{St-Onge, Iacopini,
  Latora, Barrat, Petri, Allard, and H{\'e}bert-Dufresne}}]{st2022influential}
\bibinfo{author}{\bibfnamefont{G.}~\bibnamefont{St-Onge}},
  \bibinfo{author}{\bibfnamefont{I.}~\bibnamefont{Iacopini}},
  \bibinfo{author}{\bibfnamefont{V.}~\bibnamefont{Latora}},
  \bibinfo{author}{\bibfnamefont{A.}~\bibnamefont{Barrat}},
  \bibinfo{author}{\bibfnamefont{G.}~\bibnamefont{Petri}},
  \bibinfo{author}{\bibfnamefont{A.}~\bibnamefont{Allard}}, \bibnamefont{and}
  \bibinfo{author}{\bibfnamefont{L.}~\bibnamefont{H{\'e}bert-Dufresne}},
  \bibinfo{journal}{Communications Physics} \textbf{\bibinfo{volume}{5}},
  \bibinfo{pages}{25} (\bibinfo{year}{2022}).

\bibitem[{\citenamefont{Kuramoto}(1975)}]{kuramoto1975self}
\bibinfo{author}{\bibfnamefont{Y.}~\bibnamefont{Kuramoto}}, in
  \emph{\bibinfo{booktitle}{International Symposium on Mathematical Problems in
  Theoretical Physics: January 23--29, 1975, Kyoto University, Kyoto/Japan}}
  (\bibinfo{organization}{Springer}, \bibinfo{year}{1975}), pp.
  \bibinfo{pages}{420--422}.

\bibitem[{\citenamefont{Lucas et~al.}(2020)\citenamefont{Lucas, Cencetti, and
  Battiston}}]{higher-order_kuramoto}
\bibinfo{author}{\bibfnamefont{M.}~\bibnamefont{Lucas}},
  \bibinfo{author}{\bibfnamefont{G.}~\bibnamefont{Cencetti}}, \bibnamefont{and}
  \bibinfo{author}{\bibfnamefont{F.}~\bibnamefont{Battiston}},
  \bibinfo{journal}{Phys. Rev. Res.} \textbf{\bibinfo{volume}{2}},
  \bibinfo{pages}{033410} (\bibinfo{year}{2020}).

\bibitem[{\citenamefont{D'Souza et~al.}(2019)\citenamefont{D'Souza,
  Gómez-Gardeñes, Nagler, and Arenas}}]{explosive_raissa}
\bibinfo{author}{\bibfnamefont{R.~M.} \bibnamefont{D'Souza}},
  \bibinfo{author}{\bibfnamefont{J.}~\bibnamefont{Gómez-Gardeñes}},
  \bibinfo{author}{\bibfnamefont{J.}~\bibnamefont{Nagler}}, \bibnamefont{and}
  \bibinfo{author}{\bibfnamefont{A.}~\bibnamefont{Arenas}},
  \bibinfo{journal}{Advances in Physics} \textbf{\bibinfo{volume}{68}},
  \bibinfo{pages}{123} (\bibinfo{year}{2019}).

\bibitem[{\citenamefont{Acebr{\'o}n et~al.}(2005)\citenamefont{Acebr{\'o}n,
  Bonilla, Vicente, Ritort, and Spigler}}]{acebron2005kuramoto}
\bibinfo{author}{\bibfnamefont{J.~A.} \bibnamefont{Acebr{\'o}n}},
  \bibinfo{author}{\bibfnamefont{L.~L.} \bibnamefont{Bonilla}},
  \bibinfo{author}{\bibfnamefont{C.~J.~P.} \bibnamefont{Vicente}},
  \bibinfo{author}{\bibfnamefont{F.}~\bibnamefont{Ritort}}, \bibnamefont{and}
  \bibinfo{author}{\bibfnamefont{R.}~\bibnamefont{Spigler}},
  \bibinfo{journal}{Reviews of modern physics} \textbf{\bibinfo{volume}{77}},
  \bibinfo{pages}{137} (\bibinfo{year}{2005}).

\bibitem[{\citenamefont{G\'omez-Garde\~nes
  et~al.}(2007)\citenamefont{G\'omez-Garde\~nes, Moreno, and Arenas}}]{rlink}
\bibinfo{author}{\bibfnamefont{J.}~\bibnamefont{G\'omez-Garde\~nes}},
  \bibinfo{author}{\bibfnamefont{Y.}~\bibnamefont{Moreno}}, \bibnamefont{and}
  \bibinfo{author}{\bibfnamefont{A.}~\bibnamefont{Arenas}},
  \bibinfo{journal}{Phys. Rev. Lett.} \textbf{\bibinfo{volume}{98}},
  \bibinfo{pages}{034101} (\bibinfo{year}{2007}).

\end{thebibliography}


\end{document}